# Inferring clonal composition from multiple tumor biopsies

Matteo Manica[1,2], Hyunjae Ryan Kim[3], Roland Mathis[1], Philippe Chouvarine[3], Dorothea Rutishauser[5], Laura De Vargas Roditi[5], Bence Szalai[4], Ulrich Wagner[5], Kathrin Oehl[5], Karim Saba[5], Arati Pati[3], Julio Saez-Rodriguez[4,6], Angshumoy Roy[3], Donald W. Parsons[3], Peter J. Wild[7], María Rodríguez Martínez[1], Pavel Sumazin[3]

[1] IBM Research—Zurich, 8803 Rüschlikon, Switzerland

[2] Institute of Molecular Systems Biology, ETH Zurich, Zurich, Switzerland

[3] Texas Children's Cancer Center, Baylor College of Medicine, Houston, Texas, USA

[4] RWTH Aachen University, Faculty of Medicine, Joint Research Centre for Computational Biomedicine, Aachen, Germany

[5] Pathology and Molecular Pathology, University Hospital Zurich, Zurich, Switzerland

[6] Institute for Computational Biomedicine, Heidelberg University Hospital, Heidelberg, Germany

[7] Senckenberg Institute of Pathology, University Hospital Frankfurt, Frankfurt am Main, Germany

Correspondence to Peter.Wild@kgu.de, mrm@zurich.ibm.com, and sumazin@bcm.edu

## ABSTRACT

**Background**: Knowledge about tumor clonal evolution can help interpret the function of genetic alterations by pointing out initiating events and events that contribute to the selective advantage of proliferative, metastatic, and drug-resistant tumor subclones. Clonal evolution can be reconstructed from estimates of the relative abundance (*frequency*) of subclone-specific alterations in tumor biopsies, which, in turn, inform on the tumor's composition. However, estimating these frequencies is complicated by the high genetic instability that characterizes many cancers.

**Results**: Models for genetic instability suggest that copy number alterations (*CNAs*) can influence mutation-frequency estimates and thus impede efforts to reconstruct tumor phylogenies. Our analysis suggests that accurate mutation frequency estimates require accounting for CNAs—a challenging endeavor using the genetic profile of a single tumor biopsy. Instead, we propose an optimization algorithm, *Chimæra*, to account for the effects of CNAs using profiles of multiple biopsies per tumor. Analyses of simulated data and tumor profiles suggested that Chimæra estimates are consistently more accurate than those of previously proposed methods and resulted in improved phylogeny reconstructions and subclone characterizations. We used Chimæra to infer recurrent initiating mutations in hepatocellular carcinomas, resolve the clonal composition of aggressive Wilms' tumors, and characterize the acquisition timeline of mutations that are associated with drug-resistant prostate cancers.

**Conclusions**: Explicit accounting for CNAs can dramatically improve mutation frequency estimates, leading to more accurate phylogeny reconstructions and subclone characterizations.

**Availability**: Chimæra web server and source code are available at https://ibm.biz/chimaera-aas and https://github.com/drugilsberg/chimaera. Supplementary tables are accessible from Chimæra's GitHub repository branch journal-submission.

## BACKGROUND

Pan-cancer tumor profiling has identified recurrent alterations that are associated with tumor etiology at the loci of thousands of genes but the interpretation of genetic alterations remains a major challenge [1-3]. Knowledge about the clonal evolution of tumors can point to genetic alterations that both contribute to tumorigenesis, indicate prognostically relevant intra-tumoral variability, and point to refractory tumor subclones [4, 5]. Specifically, clonal evolution—depicted as a phylogenetic tree in Figure 1A—can help identify alterations that play a role in tumor initiation as well as those that confer a selective advantage to altered tumor cells. Moreover, information about its subclone composition is important for predicting cancer's potential for drug resistance and metastasis, which vary across tumor subclones [6] and are the key determinants of patient outcome. Consequently, tumor-subclone characterization is essential for designing personalized therapies that target all tumor subclones and may hold the key to predicting tumor progression, metastases, drug sensitivity, and patient outcome.

Current methods that rely on DNA-profiling to reconstruct clonal evolution of tumors can be classified into two categories: methods that primarily rely on single-cell profiles [7-10] and those that computationally resolve mixtures of subclones from molecular profiles of bulk tumor cells, i.e., profiles of pools of cells that originate from a common malignant lesion [11-14]. Single-cell DNA sequencing can produce more definitive estimates of the proportion (*frequencies*) of tumor cells that contain each genetic alteration and more complete profiles of tumor subclones, including information about the co-occurrence of alterations within each subclone. Its primary disadvantage is operational: the availability of high-quality tumor samples that permit single-cell isolation and profiling as well as the accuracy and cost associated with parallel sequencing DNA from a multitude of cells per tumor. Moreover, improving the accuracy of single-cell mutation profiling remains challenging due to limited material availability in single cells [15]; this is not likely to improve as future sequencing technologies focus on profiling formalin-fixed paraffin-embedded (*FFPE*) tumor samples [16, 17]. Alternatively, single-cell RNA or protein profiling can help indicate tumor subclones, but these assays may not directly point to key driving genetic alterations.

Focusing on single-nucleotide somatic variants (*SNV*s; or simply mutations), we sought to reconstruct clonal evolution from DNA profiles of genetically unstable cancers. This entails deconvolving mutation frequencies, mutation-subclone associations, and CNAs from DNA profiles—including both whole-exome sequencing (*WES*) and panel-based (targeted) sequencing assays—that produce average estimates across cellular ensembles (Figure 1B-D). One approach to improve the accuracy of these deconvolutions is to profile multiple biopsies from the same tumor across time points [18] or across regions [6, 19]. This approach relies on two key assertions: (1) that genetic alterations that are specific to the same tumor subclone will co-occur with the same frequency across biopsies, and (2) that the clonal composition of heterogeneous regions varies, i.e. multiple sampling will allow for the aggregation and deconvolution of the frequencies of most mutations with improved power. It is important to note that mutations that underwent convergent evolution [20] do not violate these assertions and will not be aggregated with other mutations from the same tumor subclone because of differing frequency estimates across biopsies.

A central challenge for estimating mutation frequencies in tumors with unstable genomes is accounting for the effects of CNAs that can alter *mutated-read fractions*. These are observed in profiles of tumor biopsies that are composed of tumor cells with wild-type and mutated alleles as well as tumor-adjacent cells (Figure 1E-F). In turn, inaccurate mutation-frequency estimates can contribute to erroneous associations between mutations and tumor subclones as well as errors in phylogeny reconstructions (Figure 1G-H). We describe the mutation-frequency inference problem as that of inferring tumor subclone frequencies and associating mutations with subclones. Consequently, we describe the tumor-phylogeny reconstruction problem as that of inferring ancestral relations between tumor subclones. The main challenges for addressing the mutation-frequency inference problem are to aggregate co-occurring mutations across biopsies, estimate the frequency of each aggregate in every biopsy, and identify partial orders across aggregates that are consistent across biopsies. When viewed this way, each tumor subclone could be associated with a frequency vector that describes the proportion of cells containing its mutations in each biopsy. Ancestral order between two subclones could then be established based on (probabilistic) comparisons between their corresponding mutation frequencies. Ancestral order inference requires confident frequency assignment to the majority of mutations based on observed mutated-read fractions, and inference methods can be compared based on the number of mutations with frequency estimates, the accuracy of these estimates, and their accuracy at aggregating sister mutations that initiate the same clones.

We studied the mutation-frequency inference problem as a function of genetic instability and proposed the inference method *Chimæra* to improve these estimates and subsequent phylogeny reconstruction. Chimæra uses an optimization process to resolve the parameters of a natural model for the effects of CNAs on mutated-read fractions and is unique in its emphasis on the simultaneous inference of mutation frequencies and CNAs. We report on a comparison of Chimæra's accuracy to that of other mutation-frequency and heterogeneity inference methods on simulated DNA-profiling data of genetically unstable tumors and on biopsy subsets of thirteen tumors, including liver, kidney, and prostate cancers [21-23]. Each of these tumors was profiled in 4 to 10 regions, and 3 of the prostate cancers profiled were surveyed across multiple time points. We showed that Chimæra's inferences can be used to identify key mutations that are associated with increased subclone proliferation, drug response, and tumor grade, as well as to infer the ancestral relations between tumor subclones that harbor these mutations.

## RESULTS

We describe the results of our efforts to evaluate inference method accuracy on simulated data and to reconstruct phylogenies based on WES and targeted sequencing assays of tumor biopsies. Our analyses highlighted the challenges in inferring cancer mutation frequencies from these assays and the benefits of methods that rely on profiles of multiple biopsies per tumor to characterize tumor subclones and tumor evolution by tracking mutation aggregates.

### Simulation of DNA profiling data

We used phylogeny models—with sizes ranging from three to twelve tumor subclones, twenty to fifty somatic mutations per subclone, and varying degrees of genetic instability—to generate simulated DNA profiles. ABSOLUTE [24], AncesTree [12], EXPANDS [25], PhyloWGS [14, 26],

SCHISM [13], and Chimæra were then used to reconstruct phylogenies based on simulated data. Each method inferred ancestral relations between mutation pairs, and errors were estimated as the combined frequencies of false-positive and false-negative predictions. ABSOLUTE infers tumor purity and malignant cell ploidy directly from the analysis of somatic DNA alterations by fitting estimates of copy-ratio of both homologous chromosomes with a Gaussian mixture model, where components were centered at the discrete concentration-ratios implied by an initial frequentist estimation. AncesTree characterizes the clonal evolution of tumors based on the probabilistic model for errors in observed read fractions and infers phylogeny matrices using integer linear programming. EXPANDS clusters mutations based on their cell-frequency probability distributions; clusters are next extended by members with similar distributions and pruned based on statistical confidence by comparing the cluster maxima and peaks observed outside the core region. PhyloWGS reconstructs phylogenies based on a model for simple somatic mutations in addition to a correction for CNAs, all based on a single biopsy per tumor. SCHISM takes as input mutation cellularity estimations and mutation clustering inferred by other methods and uses a generalized likelihood ratio to infer lineage precedence and lineage divergence. A genetic algorithm is then used to build phylogenetic trees.

### Accuracy of mutation-frequency estimation based on simulated data

Our initial efforts to compare accuracy between methods based on analyses of phylogenies of size three revealed variable success rates, with some methods showing consistently poor accuracy. EXPANDS and PhyloWGS, which were designed to reconstruct phylogenies using profiles of one biopsy per tumor, and ABSOLUTE, which is best known and most effective for estimating tumor purity, had consistently poor accuracy in our simulations—with results statistically indistinguishable from random inferences. SCHISM and AncesTree had better or comparable performance than these three methods in every simulated instance. For example, the magnitude of frequency inference errors by ABSOLUTE, which processed profiles of multiple biopsies per tumor, were more than double those of SCHISM and analyses required manual parameter optimization. However, ABSOLUTE had good accuracy for inferring tumor purity in our synthetic data. SCHISM and AncesTree do not explicitly account for the full range of observed CNAs in tumors, but they were accurate in 100% of our tested instances with three tumor subclones. Consequently, we focused on accuracy comparisons between inferences by SCHISM, AncesTree, and Chimæra on phylogenies composed of 6-12 tumor subclones. Moreover, our analysis suggested that more than one biopsy per tumor was required to accurately approximate mutation frequencies and CNAs at these mutation loci.

We compared the accuracy of SCHISM, AncesTree, and Chimæra on phylogenies that were adapted from a precompiled library that was generated both manually and using CITUP [27]; see Figure 2B and Table S1 for representative phylogenies. Each somatic mutation was associated with a trio of copy numbers—$\delta^s$, $\delta_w^s$, and $\delta_m^s$ (Figure 2A)—that were taken from truncated normal distributions with means $\mu \in \{1,2,3\}$, where $\mu = 1$ corresponds to no copy number changes and standard deviation $\sigma \in \{0,1,2,3\}$; $\sigma = 0$ was used only when $\mu = 1$. The resulting copy numbers modeled a range of genetic instability conditions that were in line with observed CNAs in TCGA-profiled prostate, hepatocellular, breast carcinomas (HCC and BRCA in Figure 2D-E); we assumed no linkage between simulated CNAs of any mutations, and we omitted prostate adenocarcinoma (PRAD) curves from Figure 2 for readability. In addition, we added up to 10% of wildtype reads for all simulated mutations to account for the potential inclusion of non-tumor cells in biopsied samples (WT subclone in Figure 1A). Total coverage for each allele—i.e. the number of reads covering both the wildtype and mutated variants of a specific nucleotide—was taken by sampling mutation coverage values from prostate and Wilms' tumor biopsies profiled here. Finally, once idealized counts were available for both mutated and wildtype alleles, we simulated duplication or loss of up to 5% of the observations according to a uniform distribution. To simulate multiple regions per tumor, we repeated each biopsy simulation using the same simulation parameters but with distinct cellular composition vectors to produce simulated profiles of 6-12 biopsies per tumor, as depicted in Figure 2C. The availability of six biopsies per tumor increases the likelihood that mutations can be aggregated and subclone mutation frequencies can be compared to infer ancestral relations and are in line with pre-existing datasets, including those reported here. The selection of six biopsies is a compromise between clinical feasibility and the power needed to infer mutation frequencies and phylogenies.

AncesTree accepts no external input when estimating mutation frequencies, but SCHISM can be guided by externally inferred mutation frequencies and clusters. SCHISM's implementation includes its own selected clustering methods, and these were also used to compare accuracy. Chimæra can also be guided by externally inferred mutation frequencies and clusters, but by default, it uses a clustering approach modeled after hdbscan [28]. When comparing SCHISM and Chimæra performance on synthetic data, we clustered mutations using their native clustering approaches and with tclust [29] optimization subroutines including ElbowSSE, Entropy, GMD, Mclust, and SDIndex [30, 31]. We compared the accuracy of methods and pipelines on 2,000 simulated assays, including both with and without modeled genetic instability by varying mutation copy numbers. The accuracy of SCHISM estimates was better, on average, than that of AncesTree, but it was relatively sensitive to clustering optimization methods, with SDIndex outperforming other methods, including SCHISM's native implementation. Comparatively, Chimæra estimates were less dependent on clustering methods and significantly outperformed estimates by SCHISM with SDIndex ($p<1E-16$ by U test).

When using its native clustering approach Chimæra exhibited lower accuracy than implementations using tclust (Figure 3A), but it estimated frequencies for a significantly larger number of mutations (Figure 3B). The number of mutations with no frequency estimates by Chimæra was 2-fold less than that of the next best method, which allowed for dramatically improved phylogeny reconstruction in both simulated data and profiled cancers. Inference accuracy, for both SCHISM and Chimæra, was anti-correlated with copy number variability across biopsies. We used CNA variability as a surrogate for genetic instability and quantified it using the coefficient of variation of mutation copy numbers across biopsies, which followed truncated normal distributions (Figure 3C). However, while Chimæra inferences were affected by copy number variability, they were independent of the actual magnitude of CNAs (Figure 3D). This, in turn, suggested that instability across biopsies is a key challenge for estimating mutation frequencies. We note that both the SCHISM and tclust-based Chimæra pipelines failed to cluster 40% of mutations in our synthetic data. Moreover, while Chimæra assigned frequencies to all clustered mutations and made predictions for each simulate cancer, SCHISM did not successfully estimate mutation frequencies for some simulated genomes and cancer profiles. To account for this, accuracy comparisons in Figure 3 relied on only those mutations that had assigned frequencies by all methods. In total, our analysis suggests that mutation frequency estimation is more challenging for genomes with high copy-number variability (Figure 3C,D). All data—including supplementary tables and analyses—are available at the Chimæra GitHub repository.

### The number of regions profiled per tumor dictates algorithm convergence

Chimæra requires at least two profiled regions per tumor for making tumor subclone predictions, and while SCHISM can predict tumor subclones based on a single region, its accuracy improves with the number of profiled regions per tumor. To test the benefit of profile-multiplicity per tumor, we compared Chimæra and SCHISM analyses of multi-region WES profiles of 13 tumors using only subsets of the available tumor profiles for each analysis. Tumor profiles included profiles of 9 hepatocellular carcinomas (HCCs) [32], 3 high-risk Wilms' tumors, and a castrate-resistant prostate cancer (CRPC). Each tumor was profiled in 5-10 distinct regions, and we compared the number of subclones detected by each method in each multi-region subset as a function of the number of regions, with a minimum of two WES profiles per tumor.

All HCCs were profiled in 5 regions, which permitted testing predictions in 2-, 3-, and 4-size subsets of the assays for each tumor; Wilms' tumors were profiled in 6 and 8 regions; and the CRPC in 10 regions. Our results (Figure 4) suggested that, for most HCCs and Wilms' tumors, Chimæra-analysis of 4 tumor regions resulted in a similar number of subclones as profiling 5 tumor areas, however, profiling 2 tumor areas was insufficient for predicting tumor subclones accurately with Chimæra. Predictions for our CRPC, which had the highest genomic instability and mutation burden of the 13 tumors, largely converged with 7 profiled regions. Interestingly, because mutations that were clustered together by Chimæra were associated with sufficiently different frequencies by SCHISM, the number of subclones predicted by SCHISM were often dramatically greater, and SCHISM analyses did not produce subclone predictions for some tumors and often required more regions to converge (Figure 4B). A detailed description of tumor subclone predictions in each tumor context follows. All data and analyses are given in Table S2.

### Phylogeny inference in HCC

HCCs are high-risk liver tumors that are known to have high genetic instability [23]. We used Chimæra to infer mutation frequencies and ancestral relations between HCC subclones based on WES profiles of nine HBV-positive HCCs [32]. In total, we obtained mutated-read fractions and CNA estimates for 1,424 mutation candidates in 9 tumors and 43 tumor samples. Table S3 lists the data input to Chimæra, including mutation-frequency and CNA estimates for each mutation; it also details the outcome of the analyses described below.

Chimæra inferred frequencies estimates for 60% (858/1424) of all mutations, reconstructing phylogenetic trees for each tumor and predicting initiating clones and proliferative subclones; see representative trees in Figure 5A-C. In contrast, SCHISM inferred mutations frequency for 18% of the identified mutations. Interestingly, 100% (9/9) of the HBV-positive HCCs had predicted initiating mutations in WNT-signalling pathway genes (Figure 5D). An examination of 102 TCGA-profiled HBV-positive HCCs [23] suggested that 74% (75/102) of samples carried mutations in WNT-signalling pathway genes, and that the majority of these samples (76%) had WNT-signalling pathway mutations with mutated-read fractions above 25%—corresponding to mutations that are potentially present in the majority of cells.

To test whether WNT-signalling pathway genes were enriched for mutations—and particularly high-frequency mutations with mutated-read fractions above 25%—we calculated the proportion of tumors with such mutations in each of 186 KEGG pathways in MSigDB [33]. The most enriched pathways by p-value and mutated-sample fraction are shown in Figure 5E. P-values were estimated using permutation testing, where for each pathway, random same-size gene sets were generated using KEGG pathway genes and the mutated-sample fraction taken to generate a null distribution. WNT-singling was the most enriched pathway, and most of the remaining enriched pathways significantly overlapped it ($p<0.01$, FET). To correct for this overlap [34]—where pathways that overlap another pathway that is mutated in many samples are identified as significant—we recalculated enrichment significance for each pathway using the same test but after excluding WNT-singling pathway genes. We note that MAPK-signalling and 2 other top-10 pathways were still enriched (Figure 5E). Analyses products and input data are given in Table S3.

## Phylogeny inference in Wilms' tumors

To test SCHISM's and Chimæra's predictive ability in tumors with a range of genomic instability, we selected 3 Wilms' tumors with low- (CG118), intermediate- (CG565), and high-genomic instability (CG163). Multiple regions of these tumors were profiled by WES, including 6 regions from each of CG118 and CG163 and 8 regions from CG565 (Figure 6A). SCHISM produced stable tumor subclone predictions that agreed with predictions by Chimaera for CG118, but it did not converge on a set of subclones even when profiling was available for 7 regions from CG565; it also was not able to predict any subclones for CG163 (Figure 4B). Chimæra tumor subclone predictions converged with fewer profiled regions, and Chimæra predicted phylogenies for all 3 profiled tumors.

Chimæra analysis of CG118 profiles (Figure 6B) suggested that the tumor was composed of primarily two types of cells: tumor subclones with a *CTNNB1* (S45F) mutation and those with a mutation in *WT1* (R445W) mutation; a predicted daughter clone of the *CTNNB1* mutation had no previously studied mutations. Both *CTNNB1* and *WT1* mutations have been previously implicated with Wilms' tumor genesis [35, 36], and both clone types were present in every profiled region. However, the majority of cells—in all regions—were predicted to have *CTNNB1* mutations. To compare the effects of these mutations, we compared RNA-expression profiles in regions with the lowest and highest frequencies of *WT1*-mutated cells (Figure 6C), 7% and 3%, respectively. Analysis of the CTNNB1- and WT1-pathways [37, 38] suggested that they are differentially regulated across these tumor regions (Figure 6C). These data suggested that S45F activates CTNNB1 and that R445W inhibits WT1, as previously described [39, 40].

Chimæra analysis of CG163 profiles suggested that the acquisition of missense mutation in *LIN28A* (p.R132H) was a key event early in the formation of this tumor. Chimæra identified a second event that produced a less frequent tumor clone with many coding mutations with unknown significance. RNA-expression profiles of regions with the lowest and highest concentration of *LIN28A* mutations suggested that genes downstream of LIN28 are significantly altered in mutated cells. Analysis of CG565 did not reveal any mutations with known significance in Wilms' tumors. Data and analyses are provided in Table S4.

## Phylogeny inference in prostate tumors

To further test Chimæra's predictive ability, we studied ultra-deep profiles of multiple regions of a select set of prostate cancers at multiple time points. Ultra-deep targeted profiling allows for improved mutation identification and read-fraction estimation. A single region from each of these cancers was previously profiled and helped identify multiple predicted driver mutations [41, 42]. However, because mutations detection by WES are not always reliable and often includes both false-positive and false-negative predictions [43], we selected 3 cancers and designed a mutation panel that targets mutated genes in these cancers, as well as other known driver genes in prostate cancers [44] for ultra-deep sequencing. The identity of targeted genes is given in Methods. This approach helped test Chimæra's performance in more restrictive assays, which are quickly becoming standard in oncology clinics. Controls and 5 areas of each cancer were profiled at 2, 3, and 5 time points per cancer using both our targeted sequencing panel and OncoScan CNV Plus arrays to estimate copy number changes on genome scales; areas profiled from tumor PC1, at each of 3 time points, are shown in Figure 7A and given in Supplementary Table S5.

We recorded changes in treatment, identified mutations that may have been acquired following treatment, and predicted phylogenies for each tumor based on these mutations. We only considered mutations that were observed in multiple regions at the same time point, thus eliminating two-thirds of the candidates. Our results demonstrate the feasibility of phylogeny inference from targeted-panel profiling of multiple tumor regions and support phylogeny prediction efforts by suggesting that predicted subclones that are supported by multiple genetic variants may persist and accumulate additional genetic variants across time. We describe the results of Chimæra analysis below.

The tumor PC1 was diagnosed as Gleason 3+4 and treated with an LHRH analogue. It was profiled at 3-time points during follow-ups after treatment was started. Timepoint 1 in Figure 7A was taken over a year after diagnosis and treatment, and Time points 2 and 3 followed 1.17 and 1.4 years after Time Point 1. These suggested increased risk, with Gleason 5+5 and the discovery of a mutation in *AR* that has been associated with increased cancer cell proliferation and poor outcome. Chimæra analysis of profiles of PC1 suggested 2 predominant tumor subclones that are represented in Figure 7B by predicted deleterious mutations in *EP300* (p.I997V) and *AR* (p.T878A); Figure 7B. Tumor subclones with the *EP300* mutation were present in all regions profiled at Time Point 1. *AR* mutations were identified at high frequencies at Time Points 2 and 3 but only in regions that lacked the *EP300* mutation and had high Gleason scores. Together, these findings suggest that either the most aggressive tumor sections were not profiled in Time Point 1, or that subclones with the *AR* mutation have a proliferative advantage and have overtaken subclones with the *EP300* mutation. Protein profiling by SWATH-MS of regions with low- and high frequencies for the subclones with the *AR* mutation confirmed that genes that have been shown to be downregulated by AR [45] are significantly upregulated in *AR*-mutated regions.

The tumor PC2 was diagnosed as Gleason 5+4 and biopsied before treatment. The patient was treated for 9 months with LHRH-Analogon, during which the cancer was biopsied 2 additional times and registered an increase in severity to Gleason 5+5 at Time Point 3. Following treatment, the tumor was biopsied a 4th time and showed no change in severity (Gleason 5+5). The patient was then treated with combined androgen blockage—LHRH and Casodex—for two months, followed by a Gleason 4+5 diagnosis at Time Point 5. Time Point 1 profiles revealed the loss of *RB1* and a commonly observed stop-gain mutation in *PTEN* (p.R303X). Mutations in *BRCA2* and *EP300* were observed in Time Points 4 and 5. Chimæra analysis suggested that the RB1 locus deletion predates the *PTEN* mutation and that the BRCA2 and EP300 mutations were acquired in tumor subclones that following *RB1* loss, lacked the *PTEN* mutation (Figure 7D).

The tumor PC3 was diagnosed as Gleason 5+4 using post-treatment profiles, which included androgen blockers and orchiectomy 9 years prior to Time Point 1 profiling. A second biopsy, taken 6 months after the first suggested increased severity and Gleason 5+5. Tumor profiles identified known deleterious mutations in *PTEN* (p.Q245) and *BRCA1* (p.E1038G), as well as a stop-gain mutation in BRIP1, and nonsynonymous mutations in *BRCA2* and *PALB2*. All mutations were identified at both time points and nearly all regions. Chimæra's analysis suggested that *RB1* loss was followed by the *BRCA1* mutation. This was followed by the acquisition of a mutation in *TP53* (p.V173G) or *PALB2* and *BRIP1*. These sister clones subsequently acquired mutations in *PTEN* and *BRCA2*, respectively (Figure 7E). Comparing the expression of genes downstream from PTEN and TP53 by SWATH-MS in regions with the lowest- and highest-consternation of subclones with *PTEN* and *TP53* mutations suggested that these mutations disrupt the associated pathway (Figure 7F). Genes known to be downregulated by PTEN [46] were upregulated in regions rich with PTEN-mutated subclones as did targets of TP53 [47]. All input data—including protein expression profiles—and analyses products are given in Table S5.

## DISCUSSION

Bulk tissue DNA profiling by whole-genome and targeted sequencing can identify key DNA alterations that provide insight into the biology of tumors and indicate effective treatment options. Increasingly, these assays are used to help elucidate the clonal composition of heterogeneous tumors and even predict ancestral relationships between tumor subclones. The key to such efforts is the accurate estimation of mutation frequencies from both high- and low-coverage DNA profiles. Our study suggested that such estimation efforts must explicitly account for the copy numbers of both the reference and alternative alleles and that only accurate mutation-frequency estimates can yield accurate tumor phylogenies.

We briefly outlined current methods to reconstruct tumor clonal evolution using DNA-profiling. These include methods that rely on single-cell profiles and methods that resolve subclone mixtures from profiles of bulk tissues. We note that single-cell DNA sequencing is expected to help produce more accurate mutation-frequency estimates, but this technology is yet to mature and does not produce accurate single-cell estimates for *FFPE* tumor samples. The greatest challenge in cancer genomics—and this is not expected to change—is sample availability for patients with rich or specific clinical annotation, and FFPE is and will remain the primary preservation method for solid tumor resections. Moreover, frozen samples that could be used to generate cell suspensions that may be profiled using single-cell DNA sequencing cannot be easily used to evaluate the heterogeneity of tumors. We expect that whole-genome, whole-exome, and targeted-panel sequencing will produce the vast majority of tumor DNA profiles in the near future.

We also expect that in the cases that single-cell DNA and RNA profiling is possible and pursued, whole-genome and whole-exome sequencing will be used to inform on the biology of clones. In these cases—where analyses of single-cell RNA, DNA, or protein expression will be integrated with analyses of bulk tumor profiles—methods that impute mutation frequencies and even tumor phylogeny from bulk tumor DNA profiles will play a key role in this integrated analysis. Most importantly, we argue that given the heterogeneity of solid tumors and the often observable and documented differences in the composition of regions on the same tumor, profiling multiple tumor regions will be required for both research and clinical efforts in the future.

## CONCLUSIONS

We reported on methodology to improve the accuracy of tumor phylogeny reconstruction by improving mutation-frequency estimates from DNA profiles of multiple same-tumor biopsies. Our analysis suggested that mutation-frequency estimates are particularly challenging in the face of high genetic instability, which is characteristic of high-risk cancers, and that the accuracy of methods that rely on DNA profiles of a single biopsy of such cancers is poor. We also showed that even when profiles of multiple biopsies are available, methods that do not explicitly account for the full range of copy number variability produce inconsistent results and have poor accuracy.

Our analyses also suggested that Chimæra improves on mutation-frequency estimates by harnessing added information from multiple profiles and by directly accounting for the influence of CNAs on observations from DNA profiles. Chimæra's advantage was clearly observed in simulated data, where its performance was the most consistent and its accuracy the greatest. Interestingly, while Chimæra was able to estimate mutation frequencies with relatively high accuracy even at very high and very low copy numbers, its performance declined for the most unstable genomes where copy numbers for the same mutation varied widely across samples.

We showed that, when given a sufficient number of biopsies per tumor, Chimæra is able to address mutation-frequency estimate challenges arising from genetic instability, and that Chimæra's mutation aggregation approach can help resolve issues arising from convergent evolution [20] and false-positive mutation calls [43]. We also showed how Chimæra could be used in conjunction with ultra-deep sequencing to improve on mutation calling accuracy. In conclusion, our results suggest that accurate mutation-frequency and cellularity inference are possible using profiles of multiple biopsies per tumor when coupled with analyses that account for the effects of CNAs on observed mutation read fractions.

## METHODS

We formulated the phylogeny reconstruction problem in set-theoretic terms, which lead to a natural model for the effects of CNAs on mutated-read fractions in sequencing profiles. We describe our methodology for simulating WES tumor profiles, as well as our efforts to deconvolve mutation frequencies from simulated data using ABSOLUTE, AncesTree, EXPANDS, SCHISM, and Chimæra. Finally, to demonstrate that Chimæra can be effectively applied to clinical data, we described reconstructed phylogenies from WES profiles of ten same-tumor CRPC biopsies; a set of five same-tumor HCC biopsies from nine patients; two six same-tumor biopsies and an eight same-tumor biopsy of three Wilms' tumors; and three same-tumor prostate-cancer biopsies from three patients profiled at multiple time points.

### Phylogeny reconstruction problem

Let $M = \{m: m \in \mathbb{N}, 1 \leq m \leq n\}$ denote the set of $n$ mutations identified across a set of profiled biopsies $S$. The mutation burden in any given cell is given as a subset of $M$, $\gamma \subseteq M$, or as an element of the power set over $M$, $\mathcal{P}(M)$; i.e. $\gamma \in \mathcal{P}(M)$ is a specific mutation ensemble that characterizes a tumor subclone. We denote the cellularity of $\gamma$ and its corresponding subclone in biopsy $s \in S$ as $\rho_\gamma^s$, and the frequency of mutation $m \in \gamma$ in biopsy $s$ as $\varphi_m^s = \sum_{\{\gamma:\, \gamma \in \mathcal{P}(M),\ m \in \gamma\}} \rho_\gamma^s$. Consequently, $\sum_{\gamma \in \mathcal{P}(M)} \rho_\gamma^s = 1$ and the assignment $A = \{\rho_\gamma^s : \gamma \in \mathcal{P}(M), s \in S\}$ produces a solution to our clonality reconstruction formulation.

### Mutation frequencies

As defined above, for a mutation $m$ in biopsy $s \in S$, $\varphi_m^s$ denotes the frequency of cells in $s$ with mutation $m$. The total copy number $C^s$ of the allele targeted by the mutation can be estimated from WES data. $C^s$ is composed by: the copy numbers of the allele in cells that lack mutation $m$, $\delta^s$, the copy number of the wildtype allele in $m$-mutated cells, $\delta_w^s$ and the copy number of the mutated allele in $m$-mutated cells, $\delta_m^s$ (Figure 3). Notice that if no copy number event has occurred at the locus $m$ : $\delta^s = 2$, $\delta_w^s = 1$ and $\delta_m^s = 1$. Adopting the infinite-sites assumption, we denote the mutated-read fraction—the fraction of reads reflecting the mutated versus wild-type allele in a WES profile—in sample $s$ as $f_m^s$. Then, we can formulate the following equations (Eq. 1 and Eq. 2).

$$C^s = \delta^s\left(1 - \varphi_m^s\right) + (\delta_w^s + \delta_m^s)\varphi_m^s \quad \text{Eq. 1}$$

$$f_m^s = \frac{\varphi_m^s \delta_m^s}{C^s} \quad \text{Eq. 2}$$

Eq. 1 provides a weighted sum of the copy number contribution from each allele type, and Eq. 2 gives the ratio of the number of reads from the mutated allele and the total number of reads.

### Chimæra

Chimæra proceeds in three steps. First, mutation frequencies are approximated from sequencing and CNA data in each biopsy; then, mutations with similar frequency vectors (where each vector component gives the mutation frequency in each biopsy) are clustered together to form subclones; and finally, mutation frequencies and CNAs for these alleles are refined using an optimization process. The optimization assumes that clustered mutations that are associated with the same subclone have the same frequencies in each tumor biopsy and that $\delta_m^s$—the average copy number of $m$ (the mutated allele)—is unchanged across biopsies from the same tumor. This assumption can be relaxed in post-processing.

**A first approximation**. We first approximate the true frequency of the mutation $\varphi_m^s$ by accounting for tumor purity, i.e., the fraction of tumor cells in biopsies that include non-tumor cells, and assuming that the allele's average copy number in tumor cells—whether mutated or not—is fixed. Let $p^s$ be the purity of biopsy $s$, then Eq.2 can be rewritten as follows:

$$f_m^s = \frac{\varphi_m^s \delta_m^s p^s}{2(1-p^s)+C^s p^s} \qquad \text{Eq. 3.}$$

The experimentally observed copy number, $C_{obs}^s$, depends on the purity of the sample and the copy number of the sample tumor cells, $C^s$, as follows:

$$C_{obs}^s = 2(1-p^s) + C^s p^s \qquad \text{Eq. 4}$$

where $C_{obs}^s$ can be estimated using additional biochemical assays, genetic sequencing, or through computational analysis of WES data[48], and the normal cells are assumed to have been corrected for germline copy number variants associated biases.

The simplifying assumption that the allele's average copy number—averaged across all profiled cells—of the mutated allele in tumor cells is constant across biopsies, i.e.: $\delta_m^s = \frac{C^s}{2}$. Under this approximation, we can use Eqs. 3 and 4 to eliminate $C^s$ and obtain a first approximation of the mutation frequency $\widetilde{\varphi_m^s}$:

$$\widetilde{\varphi_m^s} = min\left(\frac{2 f_m^s C_{obs}^s}{C_{obs}^s - 2(1-p^s)}, 1\right) \qquad \text{Eq. 5.}$$

This constraint will be later removed in the optimization process that follows but is necessary at this stage to obtain a first approximation that mutation frequencies that take into account the copy number influence from WES measurements. The minimization is necessary because of the interplay between the copy numbers at these alleles that may produce a first approximation above 1.

**Subclone reconstruction**. The approximate mutation frequency vectors (Eq. 5) are next clustered to identify candidate groups of mutations that form subclones. We considered clustering algorithms with the robust treatment of outliers in order to ensure good clustering stability and quality. Specifically, we used a method modeled after *hdbscan*, a density-based hierarchical clustering method that aims at maximizing the stability of the obtained clustered against noise and requires minimal parameter selection. The number of clusters is determined automatically based on the minimal number of mutations that have to be considered to constitute a cluster. We also use tclust [29], a non-hierarchical robust clustering that trims outliers based on a probabilistic model. The number of clusters is selected by optimizing intra-cluster entropy or the sum of squared errors (SSE) and using a variety of optimization methods including the Elbow method, Gaussian mixture decomposition (GMD), and SD index[49-51]. The clustering based on *hdbscan* showed better performance on the generated synthetic data compared to others, especially when considering the number of mutations that could be assigned frequency estimates. Furthermore, it has the advantage of avoiding imposing a prior distribution on the mutations frequencies. Once the clusters are found, Chimæra assumes that each cluster represents a subclone and uses the mutation assignment to infer subclone frequencies and copy number estimates for each mutated allele in the final optimization step.

**Frequency and copy number inference**. Focusing on subclone $\gamma \in \mathcal{P}(M)$, Eq. 3 describes a relationship between the frequencies and copy numbers of mutations in $\gamma$:

$$\varphi_m^s \delta_m^s = f_m^s \frac{C_{obs}^s}{p^s} \equiv \mathcal{B}_{ms}, \qquad \forall\ m \in \gamma, s \in S. \qquad \text{Eq. 6}$$

where, $\mathcal{B}_{ms}$ is the entry of a matrix $\boldsymbol{\mathcal{B}} \in \mathbb{R}^{|S|,|\gamma|}$ corresponding to mutation $m$ and biopsy $s$. $\boldsymbol{\mathcal{B}}$ is fully determined from analysis of sequencing assays, including purity, observed copy numbers, and observed mutated-read fractions of each mutation.

Unfortunately, the right-hand side of Eq. 6 – a multiplication of frequencies and copy numbers – can not be analytically decoupled. However, from our problem formulation, mutations from the same subclone are expected to have the same mutation frequencies, i.e. $\varphi_{m_i}^s = \varphi_{m_j}^s \equiv \varphi^s\ \forall\ m_i, m_j \in \gamma$. Further, we assume that the copy number of each mutation $m$ is constant across biopsies, i.e. $\delta_m^{s_i} = \delta_m^{s_j} \equiv \delta_m \in [0, CN]\ \forall\ s_i, s_j \in S$, where $CN$ is a fixed upper bound for the copy number; $CN = 15$ in our simulations. While we expect that this assumption will introduce some errors to the approximation of $\delta_m^s$, it will have limited effects on the selection of optimal mutation frequencies because the variability of copy number averages for the mutated allele across biopsy is expected to be low for most mutated loci. We also note that we have not assumed stable genomes in our simulated data, i.e., the generated data displays variable copy numbers for the same mutated allele across biopsies in order to have an accurate estimate of the committed error.

After these assumptions, the optimization problem for each subclone $\gamma \in \mathcal{P}(M)$, based on Eq. 6, can be formulated as:

$$min\left\|\overrightarrow{\varphi^s} \otimes \overrightarrow{\delta_m} - \boldsymbol{\mathcal{B}}\right\|_2 ; 0 \le \delta_m \le CN, 0 \le \varphi^s \le 1, \forall m \in \gamma, \forall s \in S. \qquad \text{Eq. 7}$$

where $\overrightarrow{\varphi^s}$ is the mutation frequency vector across biopsies for all mutations in $\gamma$; $\overrightarrow{\delta_m}$ is the copy-number vector for each mutation in $\gamma$; $\boldsymbol{\mathcal{B}}$ is as defined in Eq. 6; and $\overrightarrow{\varphi^s} \otimes \overrightarrow{\delta_m}$ denotes the outer product of vectors $\overrightarrow{\varphi^s} \in \mathbb{R}^{|S|}$ and $\overrightarrow{\delta_m} \in \mathbb{R}^{|\gamma|}$. We used the Sequential Least-Squares Programming (SLSQP) optimization [52] to find an optimal solution of Eq. 7. To avoid being trapped in local optima, multiple runs with different random initializations for the mutations frequencies and unitary variant allele copy numbers are performed.

## Profiling and analysis of Wilms' tumors

DNA libraries were constructed and sequenced as previously described [53]. Briefly, after QC, high molecular weight double-strand genomic DNA samples are constructed into Illumina PairEnd pre-capture libraries according to the manufacturer's protocol (Illumina Inc.) with modifications. 300 base-pair fragments were checked using a 2.2 % Flash Gel DNA Cassette 5 (Lonza, Cat. No.57023). The Fragmented DNA were End-Repaired, incubated, A-tailed, and ligated. Agencourt® XP® Beads (Beckman Coulter Genomics, Inc.; Cat. No. A63882) were used to purify DNA after each enzymatic reaction. After Beads purification, PCR product quantification and size distributions were determined using the Caliper GX 1K/12K/High Sensitivity Assay Labchip (Hopkinton, MA, Cat. No. 760517). Pre-capture libraries (1µg) were hybridized in solution to VCRome 2.1 (NimbleGen) targeting 43 Mb of sequence from ~30K genes, according to the manufacturer's protocol. Sequencing was performed in paired-end mode with Illumina HiSeq 2000. Cluster formation and primer hybridization were performed on the flow cell with Illumina's cBot cluster generation system. On average, about 80-100 million successful reads, consisting of 2 X100 bp, were generated on each lane of a flow cell.

RNA-seq profiles of Wilms' tumors were aligned using STAR v2.3.0e[33] to an index of GRCh37 that included GENCODE v16 gene annotation. Alignment files were processed using Picard tools v1.54, and the final BAM files indexed using SAMtools index v0.1.11[34]. RNA-seq run quality was assessed using the RNA-SeQC package[35] using the same GENCODE 16 .gtf file. Transcript quantification was performed using Cufflinks[18] v2.02 running in quantification mode against the GENCODE v16 .gtf file. FPKM values were used for relative abundance estimation.

### Profiling and analysis of WES of prostate cancer biopsies

To test our ability to infer mutation frequencies and ancestral relations between subclones based on clinical profiles of 4 prostate cancers. The Specimen were collected at the Department of Pathology and Molecular Pathology, University Hospital Zurich, Switzerland as previously described [54] with the approval of Cantonal scientific ethics committee Zurich, approval number KEK-ZH-No. 2014-0007, and with informed consent by the patient. Tumor regions were selected for heterogeneous histological presentation by an experienced uropathologist (PJW). DNA from peripheral blood and FFPE punches was isolated with the Maxwell 16 LEV Blood DNA kit (Promega, AS1290) and Maxwell 16 FFPE Tissue LEV DNA Purification Kit (Promega AS1130), respectively, according to manufacturer's recommendations; 300µl of blood collected in a BD Vacutainer K2 (EDTA 18.0mg) tube was added to 30µl of Proteinase K solution (final concentration 2 mg/ml) and subsequently mixed with 300µl lysis buffer, vortexed and incubated for 20 min at 56°C. FFPE cylinders were deparaffinised with xylene, washed twice with ethanol, dried 10 min at 37°C and re-suspended in 200µl incubation buffer containing 2mg/ml Proteinase K. Samples were incubated overnight at 70°C and mixed with 400µl lysis buffer. Lysates from both, blood and FFPE tissues, were transferred to well 1 of the supplied cartridge of the corresponding kit and DNA was automatically purified and eluted in 30µl Tris-buffer, pH 8.0 by the Maxwell instrument.

WES samples were profiled using Agilent SureSelect Whole Exome Enrichment, v6 (58 Mbp) and 2 × 75bp paired-end reads were used for optimal performance on a HiSeq 4000 (Illumina). Mutation calling was followed by protocols established by TCGA and ExAC [21, 55]. Reads were aligned to hg19 using BWA [56], and variants were called with GenomeAnalysisTK, MuTect [57], Picard MarkDuplicates, and additional post-processing utilities from GATK including BaseRecalibrator. FastQ files were deposited in EBI's ENA project PRJEB19193. Predicted mutations are given in Table S5; mutations were annotated with estimated read fractions and estimated CNAs by VarScan using default parameters and after setting the maximum amplification to 15x [48].

## Profiling and analysis of targeted-panel sequencing of prostate cancer biopsies

We selected a total of 36 genes whose mutations are enriched in prostate cancers for ultra-deep sequencing [41, 42]. ImmQuant was used to capture their coding regions and the completeness of the capture was verified against the human reference genome GRCh38. Target genes are given below.

| | | | |
|---|---|---|---|
| AKT1 | CDH1 | MED12 | PMS2 |
| AR | CDKN1B | MRE11A | PTEN |
| ATM | CHEK2 | MSH2 | RAD51C |
| ATR | EP300 | MSH6 | RAD52D |
| AURKA | ERG | MYC | RB1 |
| BARD1 | EZH2 | MYCN | SPOP |
| BRACA1 | FOXA1 | NBN | TMPRSS |
| BRACA2 | GEN1 | PALB2 | TP53 |
| BRIP1 | HOBX13 | PIK3CA | ZNF595 |

The samples were taken from the Prostate Cancer Outcomes Cohort Study from UZH (ProCOC) and Metastatic Prostate Cancer Biobank from UZH (metaProC). Patients and target genes were selected for ultra-deep sequencing based on WES profiling and using our predictive panel. Their profiles were implemented in three batches and sequenced using Illumina HiSeq using 150bp pair-end reads by Sophia Genetics. Analysis of DNA profiles mirrored the steps described for WES analysis. BAM files are freely available on ENA project PRJEB19193.

Copy numbers were estimated using Affymetrix OncoScan arrays and the OncoScan FFPE Assay Kit for detection of genome-wide copy number changes and loss of heterozygosity in FFPE samples. Oncoscan uses molecular inversion probe technology to query over 220,000 SNPs at carefully selected genomic locations, evenly distributed across the genome and with increased density within approximately 900 cancer or cancer-related genes. All samples underwent array hybridization and analysis and passed gel QC, as established during validation of the assay for clinical genetic analysis. Data were analyzed using Nexus Express.

## SWATH-MS profiling and analysis of prostate cancer biopsies

For the comprehensive quantitation of proteins in high throughput manner, we employed a data-independent acquisition (DIA) workflow compiled of the commercially available and standardized iST sample preparation kit (PreOmics GmbH), a Q Exactive HF (Thermo Fisher Scientific Inc.) and Spectronaut Pulsar data analysis software (Biognosys AG). We have applied a combination of published and specifically generated spectral libraries. We have prepared and analyzed a selected 5 tissue samples from human prostate adenocarcinomas from the retrospective FFPE

PC cohort from the University Hospital Zurich as described above [42]. Each specimen was analyzed once.

First, the FFPE fixed material was deparaffinised and peptides were generated by using the iST 96x kit from PreOmics following the optimized protocol for FFPE punches. This was followed with DIA-MS analysis of 1ug of total peptide mixture for each patient sample including iRT standard peptides for non-linear calibration, batch correction and large scale data set merge. Then, we generated and applied spectral libraries through the merge of an in house generated library (Spectronaut Pulsar) and a commercial organ-specific library (Biognosys Nr.). The resulting library contained 8200 protein groups with a size of 65.5 mb. For quality control, the peptide yield was determined after protein extraction, digestion, and peptide clean up. MS signal intensity was monitored (TIC) during entire MS run as well as the performance and stability of the liquid chromatography according to a reference peptide set (iRT, Biognosys).

# DECLARATIONS

### Ethics approval and consent to participate

Patient samples were profiled with the approval of Cantonal scientific ethics committee Zurich, approval number KEK-ZH-No. 2014-0007, and under BCM Institutional Review Board-approved protocols.

### Availability of data and materials

BAM files are freely available on ENA project PRJEB19193. Processed (L3) data are given in supplementary tables. The Chimæra web server is available at https://ibm.biz/chimaera-aas, and source code is available at https://github.com/drugilsberg/chimaera. All supplementary tables, including synthetic data and analyses used to produce Figure 3 are also given in the GitHub repository.

### Competing interests

The authors declare no competing interests.

### Funding

This project was funded by the European Union's Horizon 2020 research and innovation programme under grant agreements 668858 and 826121 to PJW, MRM, and PS; by the Foundation for Research in Science and the Humanities at the University of Zurich to PJW, and by Texas Children's Cancer Center and Cookies for Kids' Cancer Foundation to PS and DWP.

### Authors' contributions

MM, PS, MRM, DWP, and PJW conceived the project. MM, MRM, and PS designed methods, and MM implemented methods. HRK, PC, RM, MM, BS, MRM, PS, UW, AP, JSR, and PJW analyzed data. DR, LDVR, KO, KS, AR, and PJW profiled tumors. All authors contributed to the writing of the manuscript.

## FIGURES

**Figure 1**. A simulated footprint of the clonal evolution of a tumor, as observed in genetic profiles of multiple biopsies. (**A**) Tumor phylogeny composed of six dominant tumor subclones that make up the (**B**) cellular composition of six tumor biopsies. (**C**) Read fractions in a DNA profiling assay that are associated with these subclones and (**D**) corrected fractions after accounting for CNAs. (**E**) Frequencies of the variant allele, for each mutation defining a clone in each biopsy, are higher for ancestral clones. (**F**) Variant allele frequencies are linked to the cellular composition of the tumor and depend on its associated phylogeny. (**G**) Ancestral relations can be inferred by comparing subclone frequency vectors; e.g., Subclone 3 frequencies are greater or equal to those of Subclone 4 across all biopsies, suggesting that Subclone 3 may be ancestral to Subclone 4. (**H**) However, errors in frequency estimates (red) can complicate efforts to infer ancestry and tumor-phylogeny reconstruction.

**Figure 2**. Our model for the effects of copy number alterations on mutated-read fractions in DNA profiles. (**A**) For each mutation in each biopsy $s$, the mutated-read fraction is a function of the true proportion of profiled cells with the mutation (mutation frequency) $\varphi_s$, the copy number of the reference allele in all profiled cells without the mutation $\delta_s$, and the copy numbers of the reference and the mutated allele in tumor cells with the mutation $\delta_s^0$ and $\delta_s^a$, respectively. Note that the frequency of tumor and WT cells without the mutation is $(1-\varphi_s)$ and the two reference alleles can be assumed to have a combined copy number of $2\delta_s$. (**B-E**) In order to compare inference methods we synthetically generated profiling data based on parametrized simulated copy number distributions. (**B**) Shown are representative phylogenies and (**C**) a representative cellular composition matrix, as well as (**D**) density plots of average copy numbers across profiles of TCGA-profiled hepatocellular (HCC) and breast (BRCA) carcinomas. These include the distribution of copy numbers in two individual HCCs (HCC1 and HCC2) and across all HCCs (black) and all BRCAs (blue); copy numbers ranged from 0 to >260x. (**E**) Simulated copy numbers, including simulations of more-stable and less-stable cancers, ranged from 0x to 15x copies.

**Figure 3**. Performance of mutation frequency estimates on simulated data. (**A**) Performance—measured by mean error across simulated WES datasets from genomes with varying mutation copy numbers—of mutation-frequency estimates by AncesTree (purple), SCHISM (red) and Chimæra (green and blue); SCHISM and Chimæra were evaluated using multiple clustering methods in an effort to improve their accuracy, with SDIndex (SCHISM) and ElbowSSE (Chimæra) producing top accuracy, respectively. In blue, are reported estimates for the published Chimæra, which uses *hdbscan*. (**B**) No method is able to estimate mutation frequencies for every mutation; however, Chimæra assigns frequencies for over 80% of simulated mutations, compared to an average of 60% or fewer for other methods. (**C**) Errors in frequency estimates were correlated with genetic instability, which was measured here as the coefficient of variation within copy number distributions used in simulated WES profiles. Inferences by some methods were consistently better than others; e.g., SCHISM with SDIndex clustering outperformed AncesTree inferences. Chimæra clearly outperformed all the other methods regardless of the clustering strategy. (**D**) While copy-number variability in the same

sample was correlated with inference errors, the absolute magnitude of copy numbers had no significant effect on Chimæra's performance. We report results for Chimera (hdbscan) and SCHISM with SDIndex (a representative that resembles results with other clustering methods). Standard errors are reported. Mean Error is the mean of L1 distances between true and estimated mutation frequencies after normalizing for the number of biopsies tested.

**Figure 4**. The number of predicted subclones depends on the number of profiled regions per tumor. (**A-B**) Lin et al. profiled 5 regions of each of 9 HBV-positive HCCs, and we profiled 6-8 regions of each of 3 Wilms' tumors and 10 regions of a castrate-resistant prostate cancer (CRPC). For each tumor, we exhaustively selected all subsets size 2 and up, and compared the number of predicted tumor subclones across subsets to those obtained using all available profiles using (**A**) Chimæra and (**B**) SCHISM. Our results suggested that Chimæra analysis of any 4 tumor regions resulted in a similar number of subclones as analysis of 5 tumor regions, however, profiling 2 tumor regions produced fewer predicted tumor subclones. SCHISM performed better on stable genomes (CG118) than unstable genomes (CG163 and CRPC). (**C**) Chimæra analyses (top) of subsets of HCC6046 profiles suggested convergence of subclone predictions using 3 profiled regions, while SCHISM analyses (bottom) produced a higher prediction variability across profiling subsets. (**D**) Similarly, Chimæra analyses of 4 regions of CG565 produced a similar number of clones as profiles of 8 regions. (**E**) Chimæra analyses of 7 regions of our CRPC tumor produced a similar number of clones as analysis of 10 regions; however, analyses based on 5 or fewer tumor areas produced significantly more predicted tumor subclones due to reduced aggregating power. SCHISM analyses based on 6 or more regions produced consistent counts of predicted subclones, but this number was considerably greater than the number of subclones predicted by Chimæra.

**Figure 5**. Inferred tumor phylogenies for HBV-positive HCCs suggest that WNT-signalling pathway mutations play a key role in tumor initiation. (**A-C**) Mutations in *TP53* were inferred to initiate tumorigenesis in 3 of the 9 tumors we studied; labels correspond to labeling by Lin et al. (**D**) All 9 tumors had mutations in WNT-signalling pathway genes that were predicted in the initiating tumor subclone. (**E**) The majority of the 102 TCGA-profiled HBV-positive HCCs had mutations in WNT-signalling pathway genes, which was the most significantly mutated KEGG pathway in these patients. Most other significantly mutated pathways were no longer enriched for mutations after the exclusion of WNT-signalling pathway genes from the analysis.

**Figure 6**. Inferred tumor phylogenies for high-risk Wilms' tumors identified driving initiating mutations. (**A**) The location of eight CG565 regions selection for profiling. (**B**) The inferred phylogeny for CG118 suggested that the tumor is composed of two major subclones driven by previously observed mutations in *CTNNB1* and *WT1*, with the *CTNNB1*-mutated clone accounting for a larger proportion of the tumor. CG565 was predicted to acquire mutations in *ITGA3* and *MACF1* that coincided with clonal expansion. The initiation of CG163 was predicted to include a mutation in *LIN28A*, which is sufficient to drive Wilms' tumor genesis. (**C-D**) RNA-expression profiles in (**C**) regions that were more abundant with each of the CG118 subclones—90% vs. 70%, and 30% vs. 7%, for the *CTNNB1*- and *WT1*-mutated subclones, respectively—

and (**D**) *LIN28A*-mutated subclones of CG163 (100% vs. 74%) suggested differential expression of the gene programs downstream from these predicted drivers.

**Figure 7**. Inferred phylogenies of the tumors of three prostate cancer patients. (**A**) Five regions of Prostate Cancer 1 (PC1) were profiled at each of three time points, identifying potentially deleterious mutations at each time point. (**B**) PC1's inferred phylogeny suggested that *EP300* p.I997V mutation was present at Time Point 1, and that the tumor subclone with *EP300* p.I997V (Subclone 1) is distinct from the subclone with *AR* p.T878A (Subclone 2), which was observed only in the later time point and whose clonal frequency increased with time. (**C**) Differential protein expression analysis suggested that regions that were predicted to have high clonal composition of Subclone 2 (82% vs. 0%) had higher AR-target expression than regions without Subclone 2. (**D**) Inferred phylogeny for Prostate Cancer 2 (PC2) suggested that *RB1* loss was an initiating event, and that that majority of tumor cells are the results of divergent evolution following the acquisition of *PTEN* and *BRCA2* mutations, Subclones 2 and 3, respectively. In total, 5 regions of PC2 were profiled in each of the five Time Points; Subclone 1 was observed in all time points, and Subclones 3 and 4 in Time Points 4 and 5. (**E**) PC3's inferred phylogeny included *RB1* loss as an initiating event, followed by the acquisition of a *BRCA1* mutation. These tumor cells then either acquired *TP53* and *PTEN* mutations, or *PALB2*, *BRIP1* and *BRCA2* mutations. All mutations and subclones were observed in each of the two time points profiled. The *PTEN*, *BRCA1*, and *BRCA2* mutations were previously observed in cancers. (**F**) A comparison of PC3 regions with low and high composition for Subclones 3 and 4, 90% vs. 0% and 88% vs. 0%, respectively, suggested significant differential protein expression for targets of TP53 and PTEN.

## SUPPLEMENTARY TABLES

**All supplementary tables are accessible through the Chimæra GitHub repository, branch journal-submission (https://github.com/drugilsberg/chimaera/tree/journal-submission).**

**Table S1**. Synthetic tree descriptions.

**Table S2**. Analysis of all subsets of regions in profiled tumors; associated with Figure 4.

**Table S3**. Data and Analysis of all HCC mutations.

**Table S4**. Data and Analysis of all Wilms' tumor profiles.

**Table S5**. Data and Analysis of all prostate cancer profiles.

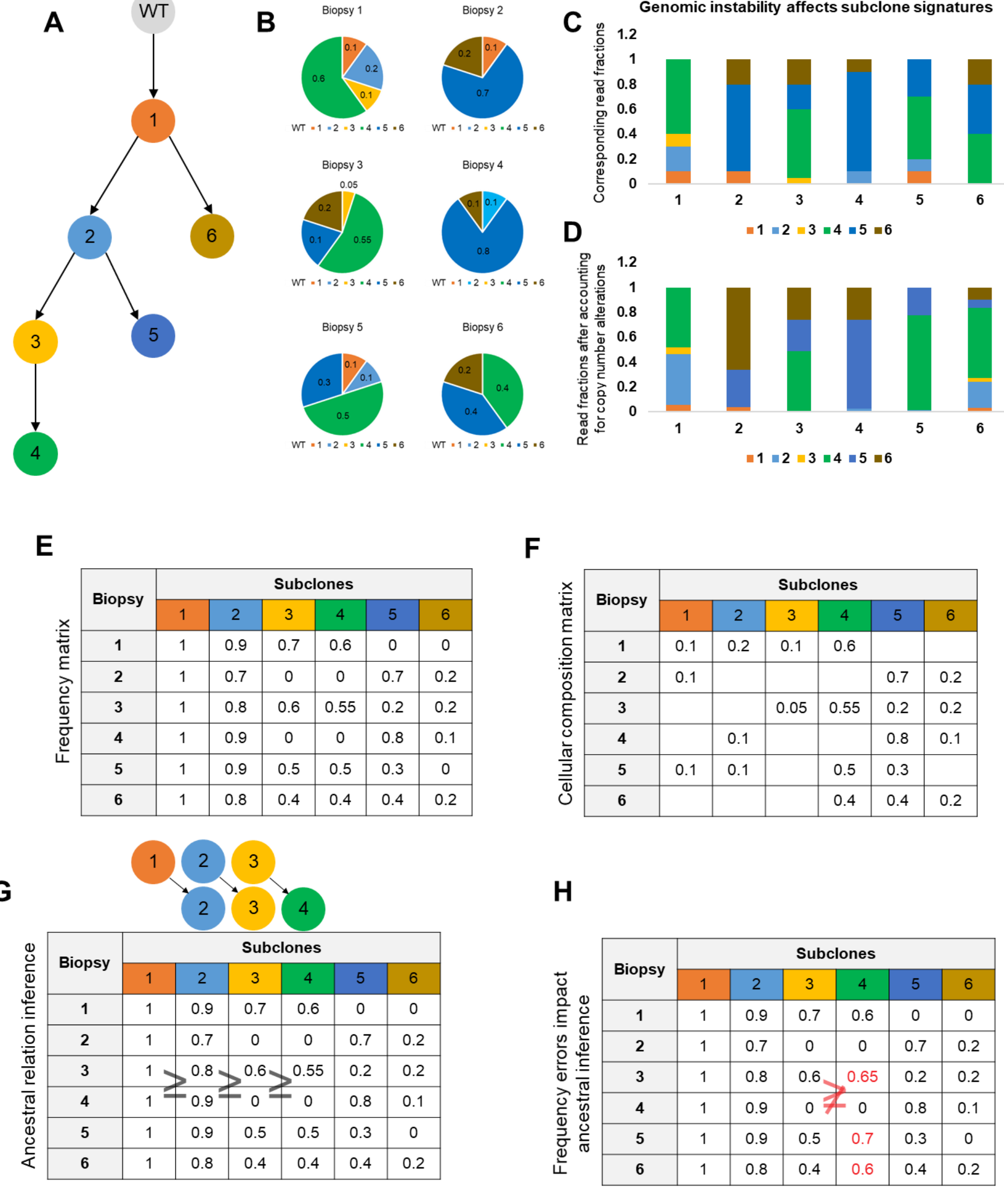

Figure 1

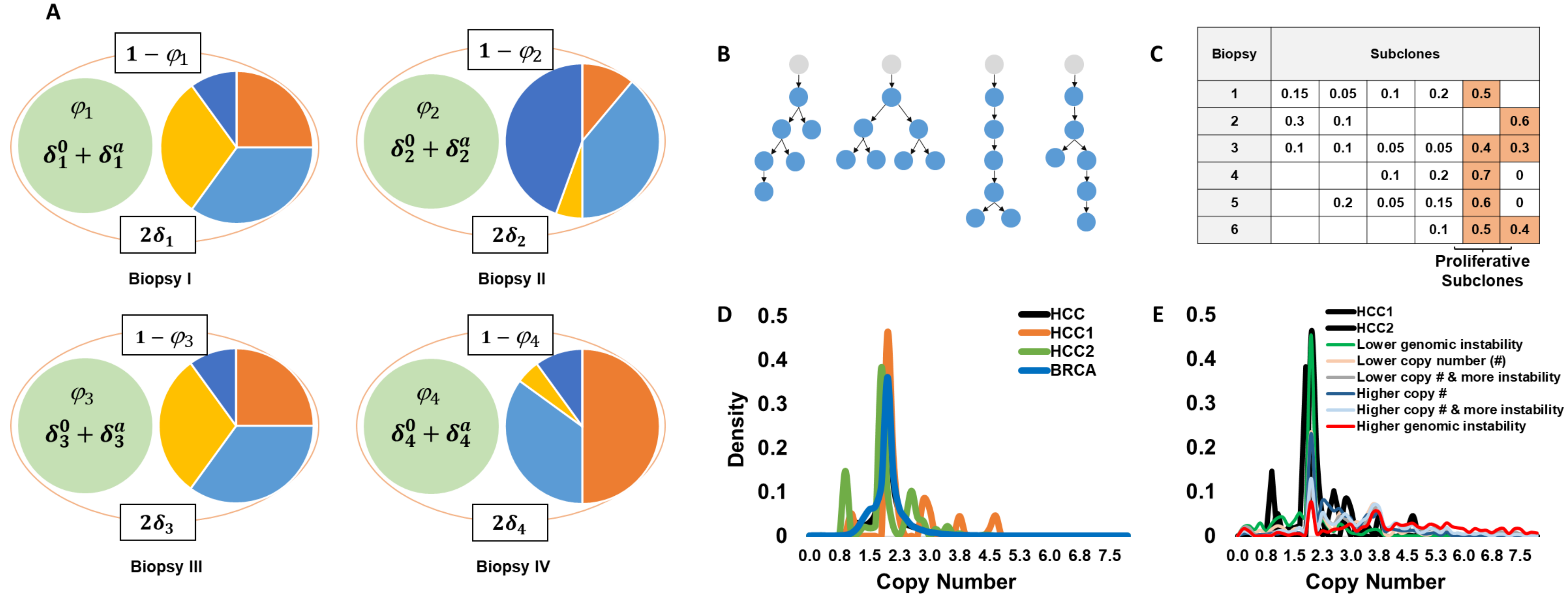

| Biopsy | Subclones | | | | | |
|---|---|---|---|---|---|---|
| 1 | 0.15 | 0.05 | 0.1 | 0.2 | 0.5 | |
| 2 | 0.3 | 0.1 | | | | 0.6 |
| 3 | 0.1 | 0.1 | 0.05 | 0.05 | 0.4 | 0.3 |
| 4 | | | 0.1 | 0.2 | 0.7 | 0 |
| 5 | | 0.2 | 0.05 | 0.15 | 0.6 | 0 |
| 6 | | | | 0.1 | 0.5 | 0.4 |

Figure 2

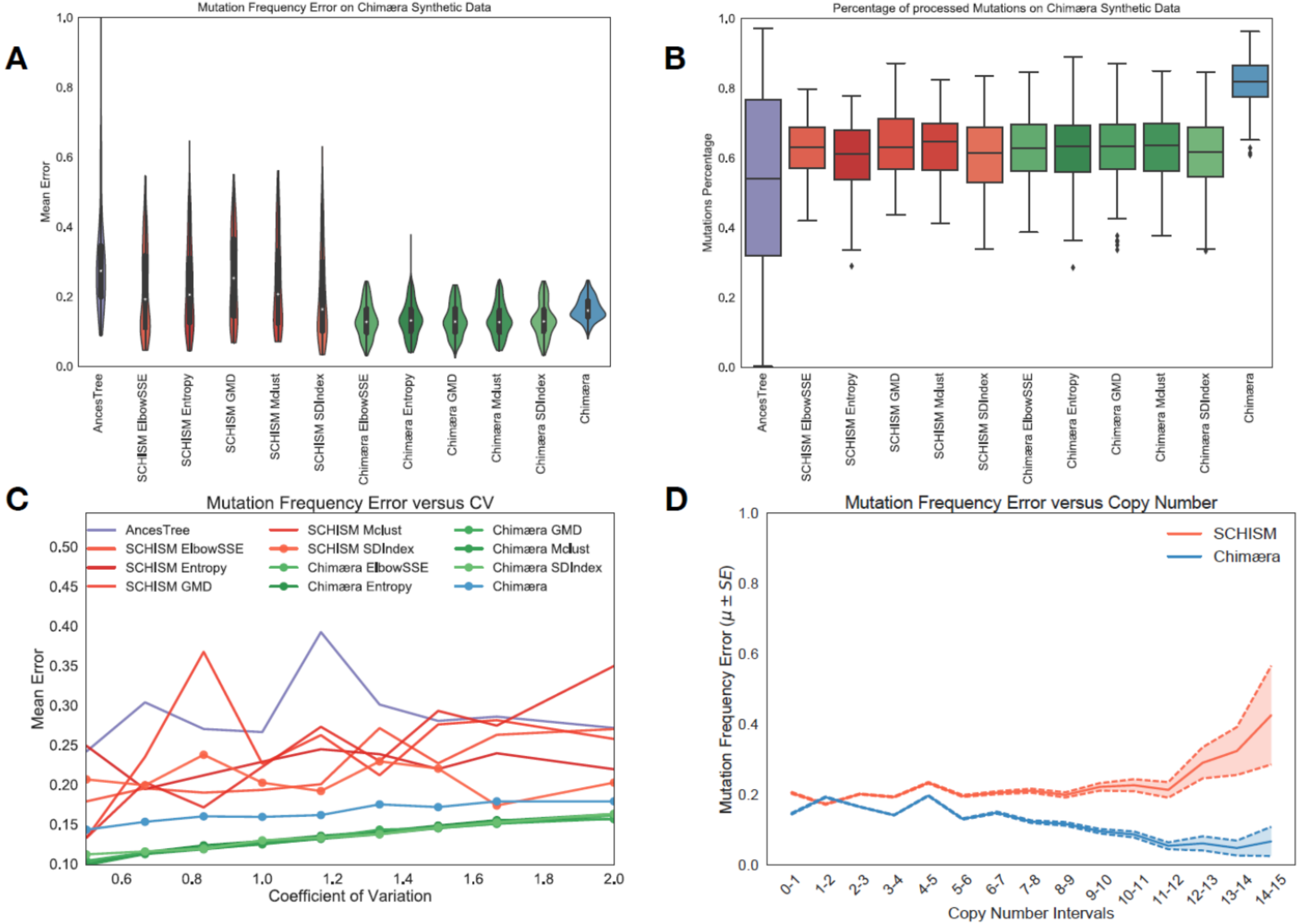

**Figure 3**

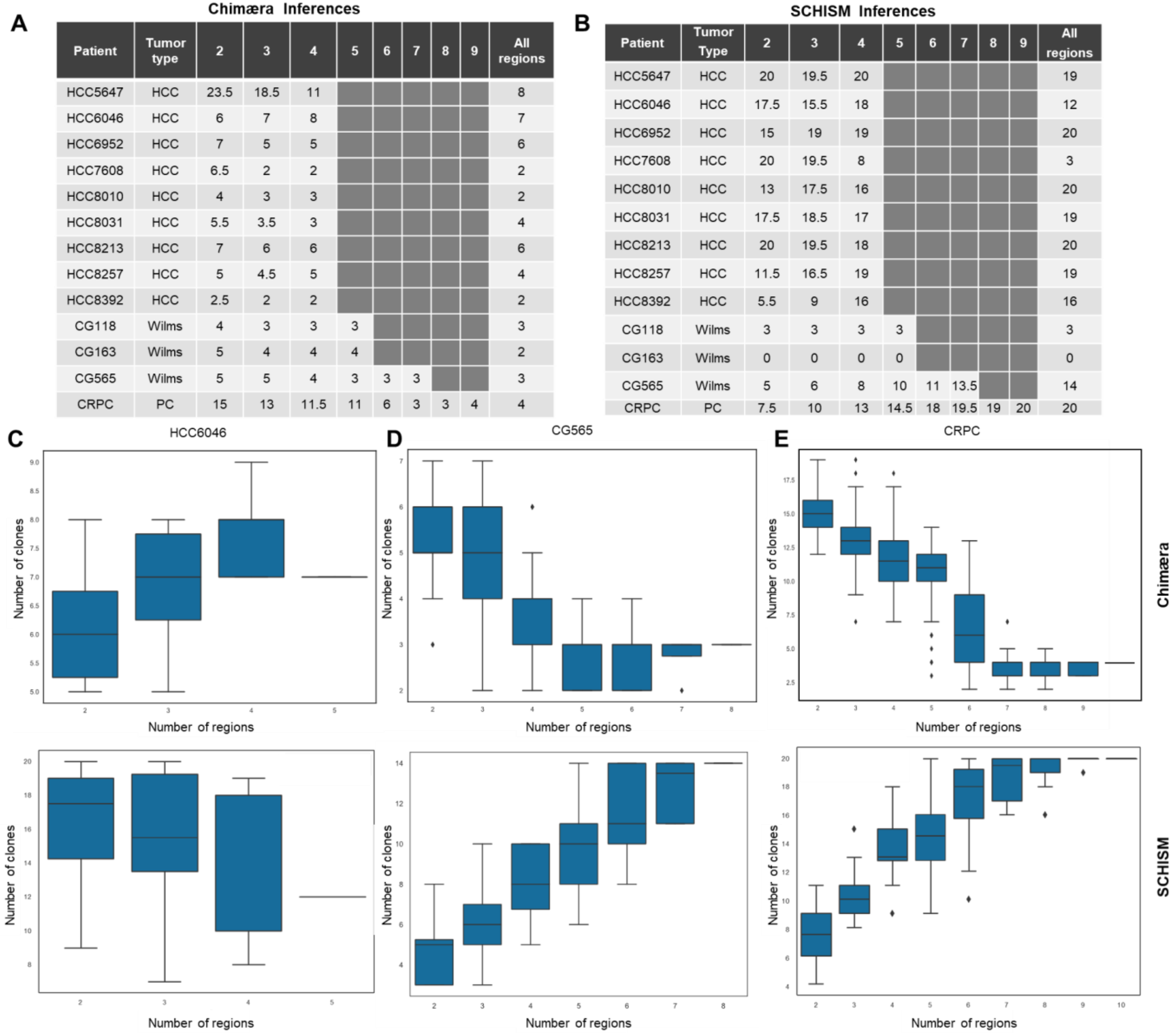

A

**Chimæra Inferences**

| Patient | Tumor type | 2 | 3 | 4 | 5 | 6 | 7 | 8 | 9 | All regions |
|---|---|---|---|---|---|---|---|---|---|---|
| HCC5647 | HCC | 23.5 | 18.5 | 11 | | | | | | 8 |
| HCC6046 | HCC | 6 | 7 | 8 | | | | | | 7 |
| HCC6952 | HCC | 7 | 5 | 5 | | | | | | 6 |
| HCC7608 | HCC | 6.5 | 2 | 2 | | | | | | 2 |
| HCC8010 | HCC | 4 | 3 | 3 | | | | | | 2 |
| HCC8031 | HCC | 5.5 | 3.5 | 3 | | | | | | 4 |
| HCC8213 | HCC | 7 | 6 | 6 | | | | | | 6 |
| HCC8257 | HCC | 5 | 4.5 | 5 | | | | | | 4 |
| HCC8392 | HCC | 2.5 | 2 | 2 | | | | | | 2 |
| CG118 | Wilms | 4 | 3 | 3 | 3 | | | | | 3 |
| CG163 | Wilms | 5 | 4 | 4 | 4 | | | | | 2 |
| CG565 | Wilms | 5 | 5 | 4 | 3 | 3 | 3 | | | 3 |
| CRPC | PC | 15 | 13 | 11.5 | 11 | 6 | 3 | 3 | 4 | 4 |

B

**SCHISM Inferences**

| Patient | Tumor Type | 2 | 3 | 4 | 5 | 6 | 7 | 8 | 9 | All regions |
|---|---|---|---|---|---|---|---|---|---|---|
| HCC5647 | HCC | 20 | 19.5 | 20 | | | | | | 19 |
| HCC6046 | HCC | 17.5 | 15.5 | 18 | | | | | | 12 |
| HCC6952 | HCC | 15 | 19 | 19 | | | | | | 20 |
| HCC7608 | HCC | 20 | 19.5 | 8 | | | | | | 3 |
| HCC8010 | HCC | 13 | 17.5 | 16 | | | | | | 20 |
| HCC8031 | HCC | 17.5 | 18.5 | 17 | | | | | | 19 |
| HCC8213 | HCC | 20 | 19.5 | 18 | | | | | | 20 |
| HCC8257 | HCC | 11.5 | 16.5 | 19 | | | | | | 19 |
| HCC8392 | HCC | 5.5 | 9 | 16 | | | | | | 16 |
| CG118 | Wilms | 3 | 3 | 3 | 3 | | | | | 3 |
| CG163 | Wilms | 0 | 0 | 0 | 0 | | | | | 0 |
| CG565 | Wilms | 5 | 6 | 8 | 10 | 11 | 13.5 | | | 14 |
| CRPC | PC | 7.5 | 10 | 13 | 14.5 | 18 | 19.5 | 19 | 20 | 20 |

**Figure 4**

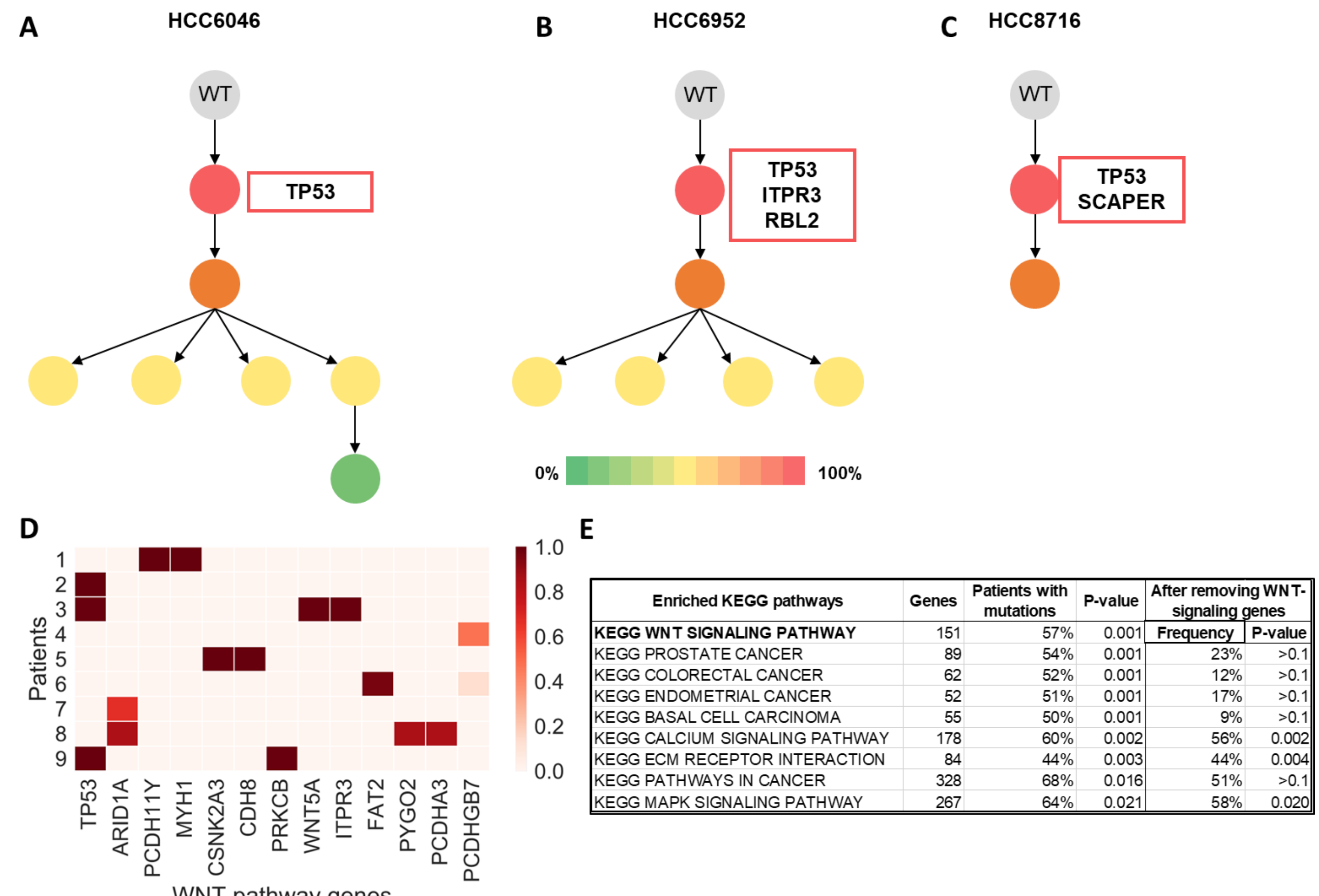

| Enriched KEGG pathways | Genes | Patients with mutations | P-value | After removing WNT-signaling genes | |
|---|---|---|---|---|---|
| **KEGG WNT SIGNALING PATHWAY** | 151 | 57% | 0.001 | Frequency | P-value |
| KEGG PROSTATE CANCER | 89 | 54% | 0.001 | 23% | >0.1 |
| KEGG COLORECTAL CANCER | 62 | 52% | 0.001 | 12% | >0.1 |
| KEGG ENDOMETRIAL CANCER | 52 | 51% | 0.001 | 17% | >0.1 |
| KEGG BASAL CELL CARCINOMA | 55 | 50% | 0.001 | 9% | >0.1 |
| KEGG CALCIUM SIGNALING PATHWAY | 178 | 60% | 0.002 | 56% | 0.002 |
| KEGG ECM RECEPTOR INTERACTION | 84 | 44% | 0.003 | 44% | 0.004 |
| KEGG PATHWAYS IN CANCER | 328 | 68% | 0.016 | 51% | >0.1 |
| KEGG MAPK SIGNALING PATHWAY | 267 | 64% | 0.021 | 58% | 0.020 |

**Figure 5**

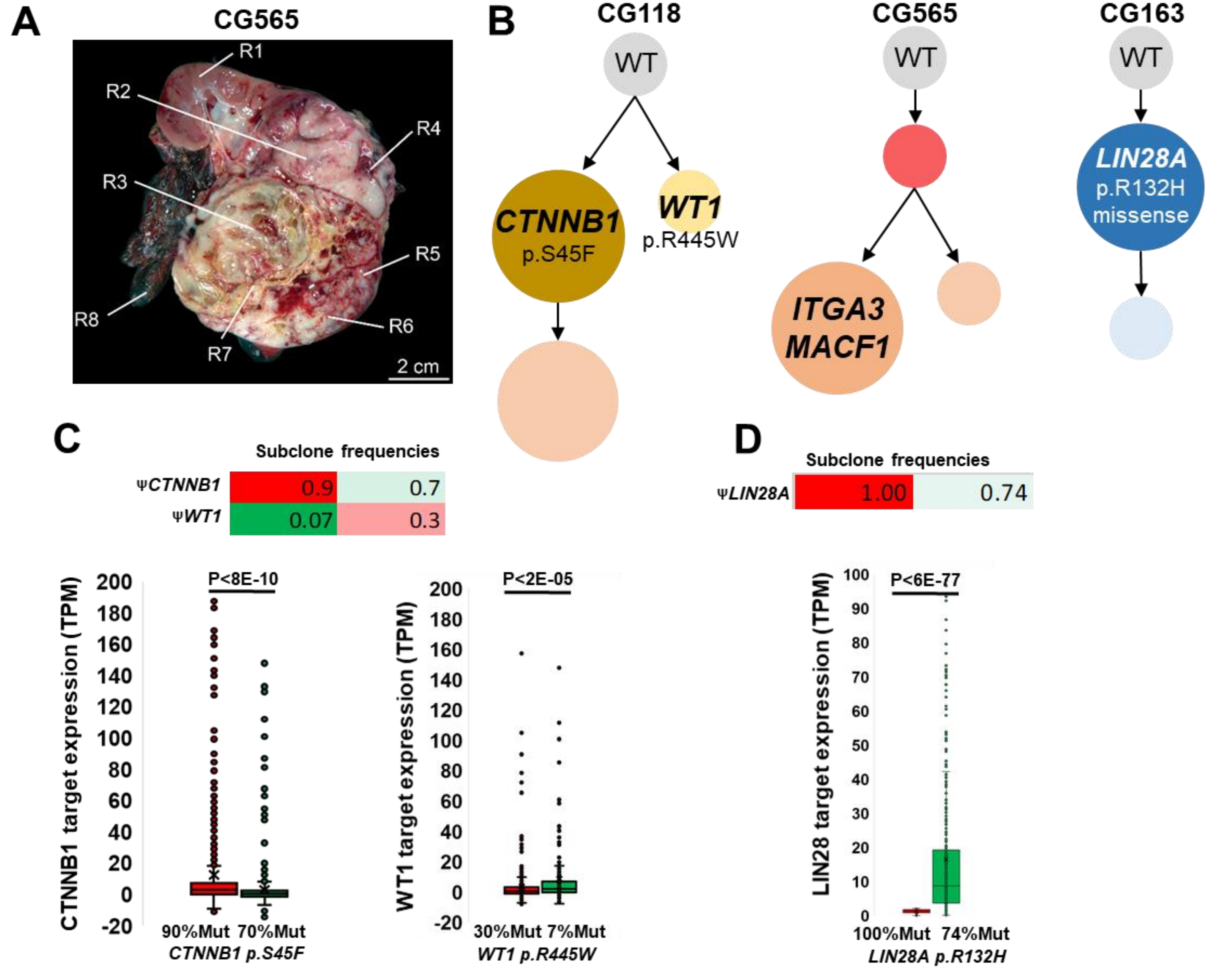

# Figure 6

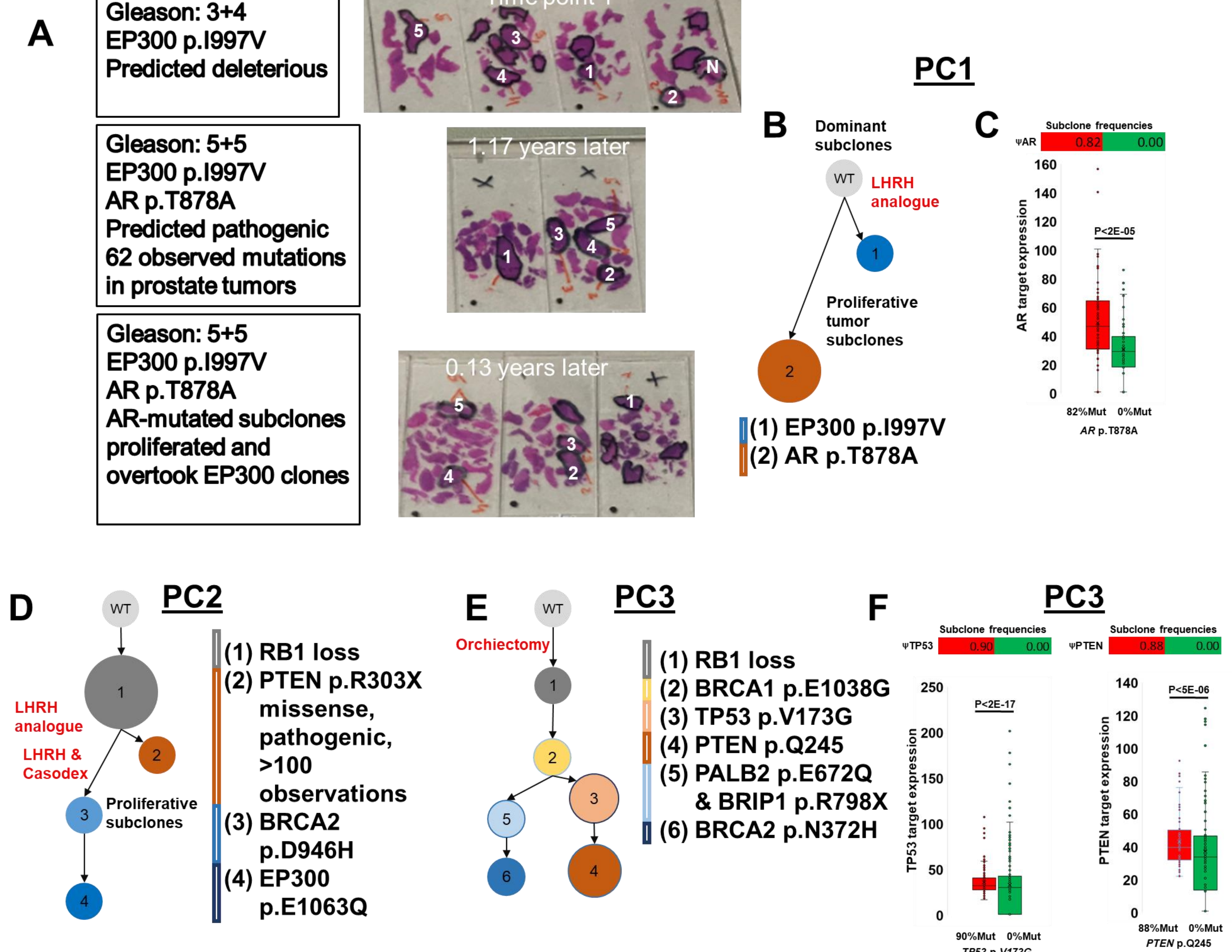

Figure 7